\documentclass{aa} 
\usepackage{twoopt}
\bibpunct{(}{)}{;}{a}{}{,}             
\makeatletter
  \newcommandtwoopt{\citeads}[3][][]{\href{http://adsabs.harvard.edu/abs/#3}%
    {\def\hyper@linkstart##1##2{}%
     \let\hyper@linkend\@empty\citealp[#1][#2]{#3}}}
  \newcommandtwoopt{\citepads}[3][][]{\href{http://adsabs.harvard.edu/abs/#3}%
    {\def\hyper@linkstart##1##2{}%
     \let\hyper@linkend\@empty\citep[#1][#2]{#3}}}
  \newcommandtwoopt{\citetads}[3][][]{\href{http://adsabs.harvard.edu/abs/#3}%
    {\def\hyper@linkstart##1##2{}%
     \let\hyper@linkend\@empty\citet[#1][#2]{#3}}}
  \newcommandtwoopt{\citeyearads}[3][][]%
    {\href{http://adsabs.harvard.edu/abs/#3}
    {\def\hyper@linkstart##1##2{}%
     \let\hyper@linkend\@empty\citeyear[#1][#2]{#3}}}
\makeatother

\makeatletter
\renewcommand*\aa@pageof{, page \thepage{} of \pageref*{LastPage}}
\makeatother

\usepackage{footnote}
\usepackage{amsmath}
\usepackage{amssymb}
\usepackage{xcolor}
\usepackage{graphicx}
\usepackage{pdflscape}
\usepackage{url}
\usepackage[varg]{txfonts}
\usepackage{bm}		
\usepackage{booktabs}
\usepackage[colorlinks=true,linkcolor=blue,citecolor=blue,urlcolor=blue]{hyperref}

\DeclareGraphicsExtensions{.pdf,.png,.jpg}

\begin{document}

\title{The first detection of dense gas in a massive main-sequence galaxy at cosmic noon}

\author{Jianhang Chen \inst{1}\thanks{ \email{jhchen@mpe.mpg.de}} 
\and
Natascha~M.~F{\"o}rster~Schreiber \inst{1} \and
Rodrigo~Herrera~Camus\inst{2,3} \and
Lilian~L.~Lee \inst{1}\and
Minju~M.~Lee \inst{4,5} \and
Claudia~Pulsoni \inst{1}\and
Daizhong~Liu \inst{6} \and
Sebastián Arriagada-Neira \inst{7} \and
Capucine~Barfety \inst{1} \and
Ric~Davies \inst{1} \and
Frank~Eisenhauer \inst{1,8} \and
Juan~Manuel~Espejo~Salcedo \inst{1} \and
Reinhard~Genzel \inst{1,9} \and
Jean-Baptiste~Jolly \inst{1} \and
Dieter~Lutz \inst{1} \and
Giovanni~Mazzolari \inst{1} \and
Stavros~Pastras \inst{1,10} \and
Alvio~Renzini \inst{11} \and
Linda~J.~Tacconi \inst{1} \and
Giulia~Tozzi \inst{1} \and
Hannah~Übler \inst{1}
}

\institute{Max-Planck-Institut f{\"u}r extraterrestrische Physik (MPE),
           Gie{\ss}enbachstra{\ss}e 1, D-85748 Garching, Germany\label{inst1}
\and
Departamento de Astronomía, Universidad de Concepción, Barrio Universitario, Concepción, Chile
\and Millenium Nucleus for Galaxies (MINGAL), Concepción, Chile 
\and Cosmic Dawn Center (DAWN), Copenhagen, Denmark
\and DTU-Space, Technical University of Denmark, Elektrovej 327, DK2800 Kgs. Lyngby, Denmark
\and Purple Mountain Observatory, Chinese Academy of Sciences, 10 Yuanhua Road, Nanjing 210023, China
\and Max-Planck-Institut f{\"u}r Radioastronomie, Auf dem Huegel 69, D-53121 Bonn, Germany
\and Technical University of Munich, TUM School of Natural Sciences, Physics Department, 85747 Garching, Germany
\and Departments of Physics and Astronomy, University of California, Berkeley, CA 94720, USA
\and Max-Planck-Institut für Astrophysik (MPA), Karl-Schwarzschild-Str. 1, D-85748 Garching, Germany
\and INAF - Osservatorio Astronomico di Padova, Vicolo dell’Osservatorio 5, Padova I-35122, Italy
          }

\date{Received 2026 ** * / Accepted **** ** *}

\titlerunning{dense gas in BX610}
\authorrunning{Jianhang Chen et al.}

\abstract{

Dense gas is the direct fuel for star formation, but measuring it has long been difficult at $z\ge2$, especially in typical star-forming main-sequence galaxies.
In this work, we report the first detection of HNC (J = 5--4) and CN (N = 4--3) emission in a massive main-sequence galaxy, BX610, at $z=2.21$.
The velocity integrated emission of HNC(5--4)+CN(4--3) is concentrated in the galactic centre, coincident with the region of ongoing intense star formation.
Based on line decomposition, we measure a line flux ratio HNC(5--4)/CN(4--3) of $1.05\pm0.23$, similar to that of starburst galaxies at comparable redshifts but lower than that of quasar/AGN host galaxies.
The comparatively fainter HNC(5--4) disfavours the presence of a strongly buried AGN in BX610, consistent with optical line diagnostics.
The radiative transfer analysis favours the presence of dense gas with a density of $(2-4)\times10^{6}\,\text{cm}^{-3}$ and a kinetic temperature of $50$--$80\,\text{K}$.
The derived abundance ratio between $N$(HNC) and $N$(CN) favours dense gas clouds near photodissociation regions, as commonly seen in typical starburst environments.
The inferred dense-gas line luminosity closely follows the scaling relation between far-IR and dense-gas line luminosities established for local luminous infrared galaxies (LIRGs).
Our observations support the view that star formation in cosmic noon galaxies is primarily controlled by the availability of dense gas, which could be enhanced in central galactic regions with efficient cold gas inflows as observed in BX610 along the inner spiral arms and a possible stellar bar.
}


\maketitle
\nolinenumbers

\section{Introduction}
\label{sec:introduction} 

Star formation, a fundamental process driving galaxy evolution, takes place within the cold, dense cores of giant molecular clouds (GMCs).
Understanding the mechanisms that regulate the conversion of gas into stars across cosmic history requires a detailed inventory of the molecular gas reservoirs that fuel star formation.
For decades, the rotational transitions of carbon monoxide (CO) have been the workhorse for tracing the bulk molecular gas content of galaxies \citep[see reviews][]{Solomon2005,Carilli2013,Tacconi2020,Saintonge2022,Schinnerer2024}.
However, the widely used low-J transitions are easily excited and can be optically thick; their emission therefore arises mainly from the surfaces of molecular clouds and traces the widespread, lower-density molecular gas rather than directly probing the compact, high-density regions where star formation actually occurs \citep{Gao2004}.

To trace the gas that immediately precedes star formation, it is necessary to observe molecules that require higher gas densities ($n({\rm H}_2) > 10^4\,\text{cm}^{-3}$) to be collisionally excited \citep{Gao2004}.
Molecules with high dipole moments, such as hydrogen cyanide (HCN) and its isomer hydrogen isocyanide (HNC), are therefore better probes of the dense molecular gas.
Indeed, observations in the local Universe have revealed a remarkably tight, linear correlation between the luminosity of the HCN(1--0) line (tracing the total dense molecular gas mass) and the far-infrared (FIR) luminosity (tracing the total star formation rate) \citep[][]{Gao2004}.
This linear relationship, $L({\rm FIR}) \propto L({\rm HCN})$, holds across a wide range of physical scales, from giant molecular clouds to the most luminous starburst galaxies \citep[e.g.][]{Gao2004a,Wu2005,Lada2012,Garcia-Burillo2012,Kauffmann2017,Jimenez-Donaire2019,Dame2023,Neumann2023}.
Similar linear correlations have also been found for other dense gas tracers, including those with higher critical densities ($n_\text{crit}$) \citep[e.g.][]{Wang2011,Zhang2014,Tan2018,Nishimura2024}, which further supports the dense gas mass as the primary driver of star formation.
In contrast, the relationship between FIR and total molecular gas luminosity is non-linear and steepens for more luminous galaxies, indicating that as the star formation rate increases, a larger fraction of the total molecular gas is in a dense state \citep{Gao2004,Lada2012}.

Extending these studies to ``cosmic noon'' (redshift $z\,\sim\,1-3$), the peak epoch of cosmic star formation \citep{Madau2014}, is critical for understanding galaxy growth in general.
However, observing dense gas tracers at these distances is challenging because of their intrinsic faintness \citep{Bethermin2018,Oteo2017,Rybak2022}.
Consequently, most high-$z$ detections of dense gas have targeted the most extreme and luminous galaxy populations, such as submillimeter galaxies (SMGs), quasars, and the more extreme hyperluminous infrared galaxies (HyLIRGs), and have often relied on gravitational lensing to boost the signal \citep{Carilli2005,Danielson2011,Spilker2014,Oteo2017,Bethermin2018,Canameras2021,Yang2023,Rybak2022,Rybak2026,Bakx2026}.
These initial studies have hinted that the properties of dense gas in high-$z$ starbursts may differ from those in the local Universe, with some evidence suggesting a potential deviation from the local linear FIR--HCN correlation \citep{Gao2007,Bethermin2018,Rybak2022}.

While these observations of extreme populations have been invaluable, they may not be representative of the broader galaxy population.
A significant gap remains in our understanding of the dense gas content of more typical, ``main-sequence'' star-forming galaxies, which dominate the population and cosmic star formation rate density at cosmic noon \citep[see recent reviews by][]{Tacconi2020,ForsterSchreiber2020}.
Understanding the dense gas properties of these more common systems is essential for a complete picture of galaxy evolution during this era.

In this letter, we report the first detection of HNC J=5--4 (hereafter HNC(5--4)) and CN N=4--3 (hereafter CN(4--3)) in a massive star-forming main-sequence galaxy at $z=2.21$.
The letter is organised as follows.
We summarise the observations, data, and analysis in \S\ref{sec:data} and present the results in \S\ref{sec:results}.
We discuss the results and present our conclusions in \S\ref{sec:discussion}.
Throughout the paper, we adopt a $\Lambda$CDM cosmology with $H_0 = 70$~km\,s$^{-1}$\,Mpc$^{-1}$ and $\Omega_\text{m} = 0.3$.

\begin{table}[th]
  \centering
  \caption{Details and measurements on BX610}
  \label{tab:measurements}
  \begin{tabular}{cccc}
    \hline\hline
    Parameter & Values & Units \\
    \hline
    R.A.${}^a$ & 23:46:09.4 & h:m:s \\
    Dec.${}^a$ & 12:49:19.2 & d:m:s \\
    log($L_\text{IR}/\text{L}_\odot$)${}^b$ & 12.63$\pm$ 0.1& -- \\
    log(SFR/(M$_\odot \text{yr}^{-1}$))${}^b$ & 2.15$\pm$ 0.1& -- \\
    $I_\text{CO(4--3)}$ & 1.73$\pm$0.07 & Jy\,km\,s$^{-1}$ \\
    $I_\text{CO(4--3), centre}$ & 0.36$\pm$0.02 & Jy\,km\,s$^{-1}$ \\
    FWHM$_\text{CO(4-3)}$ & 294$\pm$49 & km\,s$^{-1}$ \\
    FWHM$_\text{CO(4-3), center}$ & 236$\pm$4 & km\,s$^{-1}$ \\
    $I_\text{HNC+CN}$$^{c}$ & 50$\pm$20 & mJy\,km\,s$^{-1}$ \\
    FWHM$_\text{HNC+CN}$ & 300$\pm$30 & km\,s$^{-1}$ \\
    $I_\text{HNC+CN, centre}$ & 27$\pm$3 &  mJy\,km\,s$^{-1}$ \\
    $I_\text{HNC(5--4), centre}$ & 13.8$\pm$4.0 &  mJy\,km\,s$^{-1}$ \\
    FWHM$_\text{HNC(5-4)}$ & 210$\pm$60 & km\,s$^{-1}$ \\
    $I_\text{CN(4--3), centre}$ & 13.2$\pm$6.4 & mJy\,km\,s$^{-1}$ \\
    FWHM$_\text{CN(4-3)}$ & 210$\pm$60 & km\,s$^{-1}$ \\

    $L^\prime_\text{HNC+CN}/L^\prime_\text{CO(4--3)}$ & 0.03$\pm$0.01 & -- \\
    $L^\prime_\text{HNC+CN, centre}/L^\prime_\text{CO(4--3), centre}$ & 0.08$\pm$0.01 & -- \\
    $L^\prime_\text{HNC(5--4), centre}/L^\prime_\text{CN(4--3), centre}$ & 1.05$\pm$0.23 & -- \\
    \hline
  \end{tabular}
\raggedright\small{\textbf{Notes:} \\
$^{a}$ \cite{Erb2004}, $^{b}$ \cite{Brisbin2019}.\\
$^{c}$ HNC+CN is an abbreviation for HNC(5--4)+CN(4--3).}
\\
\end{table}

\section{Data and Analysis}
\label{sec:data}

\begin{figure*}[t]
  \centering
  \includegraphics[width=0.79\textwidth]{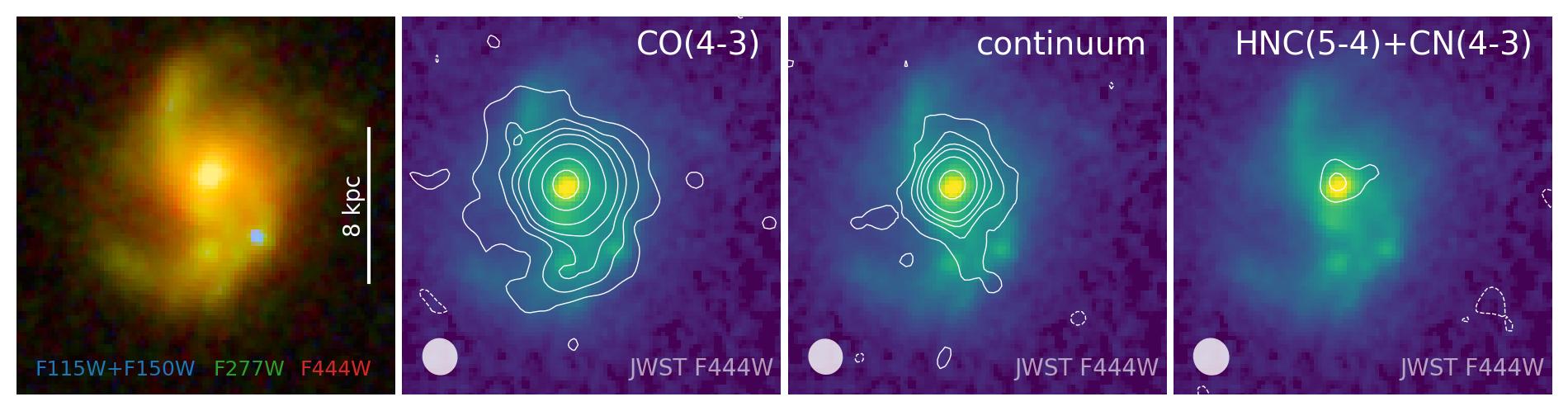}
  \includegraphics[width=0.19\textwidth]{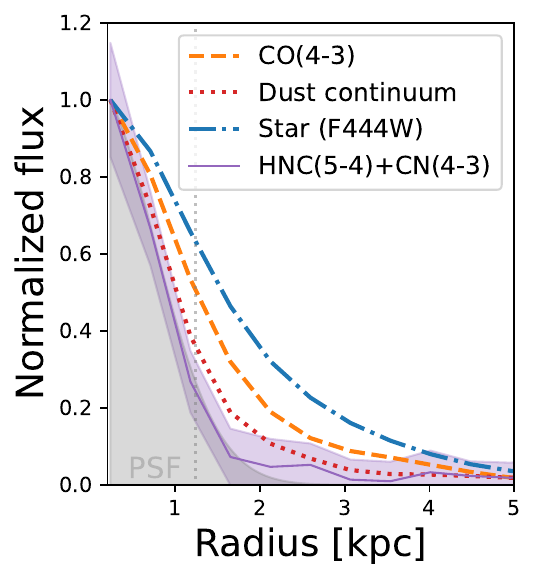}
  \caption{The colour image of BX610 and the spatial comparison between the stellar emission (\textit{JWST} F444W) and the molecular gas distributions traced by CO(4--3), the dust continuum at rest-frame 650\,$\mu$m, and the dense gas traced by HNC(5--4)+CN(4--3). The colour image is constructed from \textit{JWST} broadband images, with the colour channels indicated by the text. The white contours in the three middle panels show the line or dust emissions, starting at $3\sigma$ of the noise level and increasing in steps of $3\sigma$ for CO(4--3) and dust ($1\sigma$ for HNC(5--4)+CN(4--3)). The last panel show the radial profiles of all the different components along the major axis. The grey shadow shows the PSF of the HNC+CN image. HNC+CN emission is concentrated in the galactic centre.}
  \label{fig:maps}
\end{figure*}

Q2343-BX610 (hereafter BX610), a bright main-sequence galaxy at $z=2.21$, has been extensively observed at optical and near-infrared wavelengths \citep{Erb2004,Steidel2004,ForsterSchreiber2006,Genzel2008,ForsterSchreiber2009,ForsterSchreiber2011,ForsterSchreiber2014,ForsterSchreiber2018} as well as in the far-infrared and sub-millimetre \citep{Tacconi2013, Aravena2014,Bolatto2015,Brisbin2019,Arriagada-Neira2025}.
It exhibits two prominent spiral arms embedded in a rotation-dominated disk, characterised both by ionised gas \citep{ForsterSchreiber2011} and by cold molecular gas \citep{Genzel2023}.
It is an active star-forming galaxy, with a total molecular gas mass of $1.2\times10^{11}\,\text{M}_\odot$ \citep{Arriagada-Neira2025} and an SFR of about $141\pm6\,\text{M}_\odot\,{\rm yr}^{-1}$ \citep{Brisbin2019}.
We summarise the integrated properties of BX610 in Table~\ref{tab:measurements}.

\subsection{ALMA observations}

The wavelength range of the HNC(5--4) and CN(4--3) emission lines has been covered by programs targeting the CO(4--3) and [C\,\textsc{i}](1-0) emissions: 2013.1.00059.S (P.I. M. Aravena), 2017.1.01045.S (P.I. D. Brisbin), and 2019.1.01362.S (P.I. R. Herrera-Camus), totaling an on-source time of 19.2 hours (see also \citealt{Arriagada-Neira2025} and \citealt{Lee2024}). 
For project 2013.1.00059.S, we retrieved the data from the Additional Representative Images for Legacy (ARI-L) database\footnote{https://sites.google.com/inaf.it/ari-l/products}, which systematically recalibrated the ALMA Cycle~2--4 archive data using the more recent, mature calibration pipelines.
For projects 2017.1.01045.S and 2019.1.01362.S, we used the official pipeline-calibrated data shipped with the data delivery.
To facilitate data combination, we regridded the visibilities from the different observations to the same reference frame as the most recent observation, 2019.1.01362.S, using CASA task \textsc{mstransform} shipped with the CASA pipeline \citep[ver.~6.5.4,][]{CASATeam2022}.

We first subtracted the continuum emission directly from the visibilities using the CASA task \textsc{uvcontsub}. 
The continuum channels were determined by excluding all possible spectral lines using the dirty datacube.
We then recalculated the visibility weights for each observation with the CASA task \textsc{statwt}.
The combined visibilities were then imaged with \textsc{tclean}.
We adopted natural weighting to achieve the best sensitivity to the faint HNC(5--4) emission.
The imaging was cleaned down to the $1\sigma$ level.
We also re-imaged the CO(4--3) data following the same scheme.
For the spectral line intensity map, we collapsed all spectral line channels selected by visual inspection.
The combined datasets deliver the deepest CO(4--3) and dust emission maps with kpc-scale resolution (see Fig.~\ref{fig:maps}).

\subsection{JWST observations}

BX610 has been observed by JWST/NIRCam with the broad-band filters F115W, F150W, F277W, and F444W (GO 9407. PI: Rodrigo Herrera-Camus).
The data were calibrated using a customised JWST pipeline\footnote{https://github.com/1054/Crab.Toolkit.JWST} which followed the COSMOS-Web imaging strategy \citep[][see also \citealt{EspejoSalcedo2025}]{Franco2026}.
The RGB colour image is shown in Fig.~\ref{fig:maps}.


\subsection{Spectral line measurements}

\begin{figure*}[htpb]
  \centering
  \includegraphics[width=0.95\textwidth]{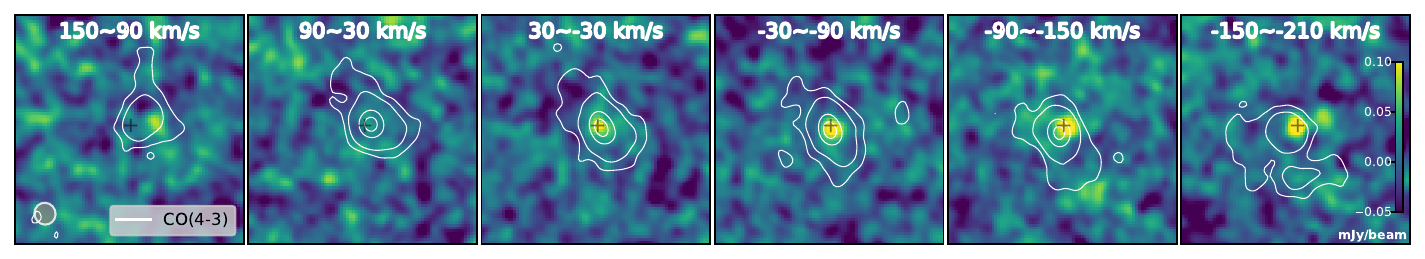}
  \includegraphics[width=0.95\textwidth]{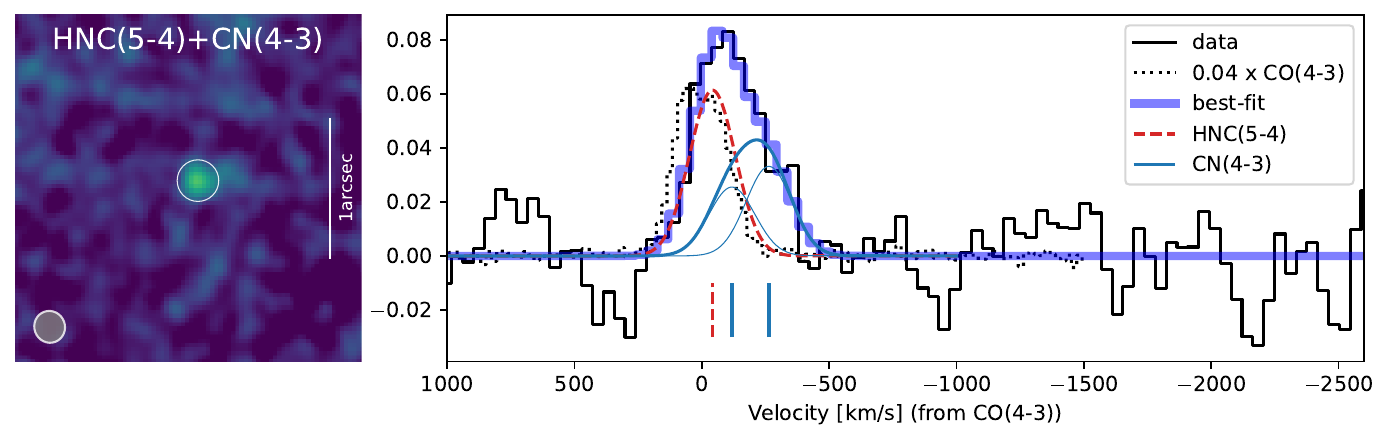}
  \caption{Spectral line decomposition of the HNC(5--4)+CN(4--3) line emission and its comparison with CO(4--3). \textit{Top:} channel maps of HNC(5--4)+CN(4--3) at different velocity ranges; the white contours show the emission from CO(4--3). \textit{Bottom:} spectral line decomposition of HNC(5--4)+CN(4--3) extracted from the galactic centre. The extraction aperture is shown as the white circle, yielding the highest total S/N. The line decomposition indicates comparable contributions from HNC(5--4) and CN(4--3) to the total emission. The extracted CO(4--3) spectrum from the same region is shown as the black and dotted line (scaled by a factor of 0.04). The vertical short lines show the best-fit velocity centre of HNC(5--4) and CN(4--3), relative to CO(4--3), which show a small velocity offset $\sim37\pm26\,\text{km\,s}^{-1}$.}
  \label{fig:spectra}
\end{figure*}

We measured the line flux in two ways, focusing on both the global integrated value and the value measured from the galactic centre.
Since HNC(5--4) and CN(4--3) are blended, we treat them jointly for the total flux measurement.
We first measured the total flux from the spectral line maps using the curve-of-growth method.
We gradually increased the aperture size, using the shape of the cleaned beam to extract the flux, and quoted the integrated flux as the value at which the curve of growth first reaches a maximum.
As shown in Fig.~\ref{fig:maps}, HNC(5--4)+CN(4--3) emission comes mainly from the centre.
We thus extracted the spectrum with a circular aperture with $r$=0.15" ($\sim1.2$\,kpc), centred on the HNC(5--4)+CN(4--3) emission.
This extraction yields the highest total signal-to-noise ratio (S/N) for the spectrum, which is key for the subsequent line decomposition.
The aperture is illustrated as a white circle in Fig.~\ref{fig:spectra}.
We fitted the spectrum (see next paragraph) to get the total flux.
To better compare its line ratio with that of CO(4--3), we also extracted the CO(4--3) spectrum with the same aperture.
We refer to this second measurement as the central flux. 
We also applied the same aperture to derive the central IR luminosity $L_\text{IR}$, by scaling the total IR luminosity according to the central dust luminosity fraction.

We decompose the contributions from HNC(5--4) and CN(4--3) with the spectrum extracted from the galactic centre.
HNC(5--4) is a single-component line with a rest-frame frequency of 453.27~GHz.
CN has a complex energy-level structure owing to its spin-coupling and spin--spin interactions.
In the N=4--3 transition, its line emission is concentrated around the rest-frame frequencies of 453.39 and 453.61~GHz, with an intensity ratio of about 1:1.3 \citep{Guelin2007}.
We simultaneously fit the two lines with three Gaussian profiles: one for HNC(5--4) and two for CN(4--3).
We fixed the intensity ratio between the two components of CN(4--3) to the theoretical value of 1.3.
We also fixed the velocity separation between CN(4--3) and HNC(5--4) and set their velocity dispersions equal.
Fig.~\ref{fig:spectra} shows the best fit, and the values are reported in Table~\ref{tab:measurements}.
The best-fit line ratio between HNC(5--4) and CN(4--3) is $1.05\pm0.23$.
The joint fit yields a better result than fits including only HNC(5--4) or only CN(4--3).
Moreover, fitting only HNC(5--4) or only CN(4--3) implies a large velocity offset ($>100$~km\,s$^{-1}$) relative to the CO(4--3) extracted from the same region.
We therefore conclude that the detections of both HNC(5--4) and CN(4--3) are robust.

\section{Results}
\label{sec:results}

\begin{figure*}[htpb]
  \centering
  \includegraphics[width=0.32\textwidth]{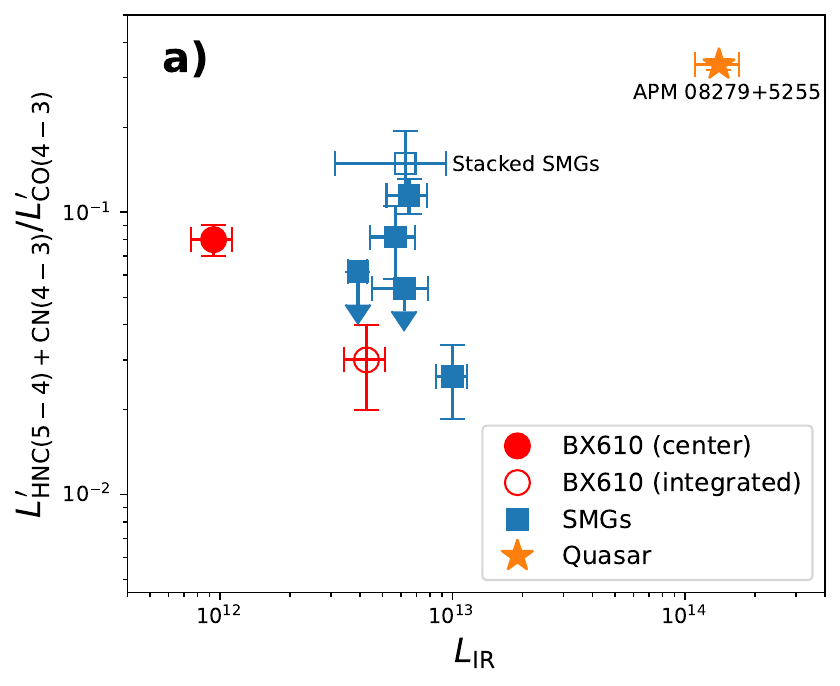}
  \includegraphics[width=0.32\textwidth]{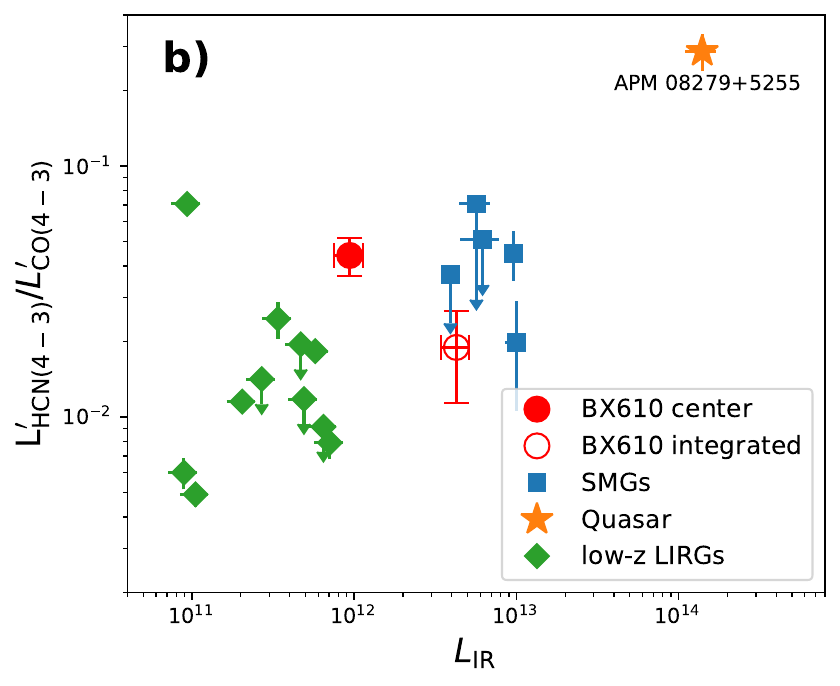}
  \includegraphics[width=0.31\textwidth]{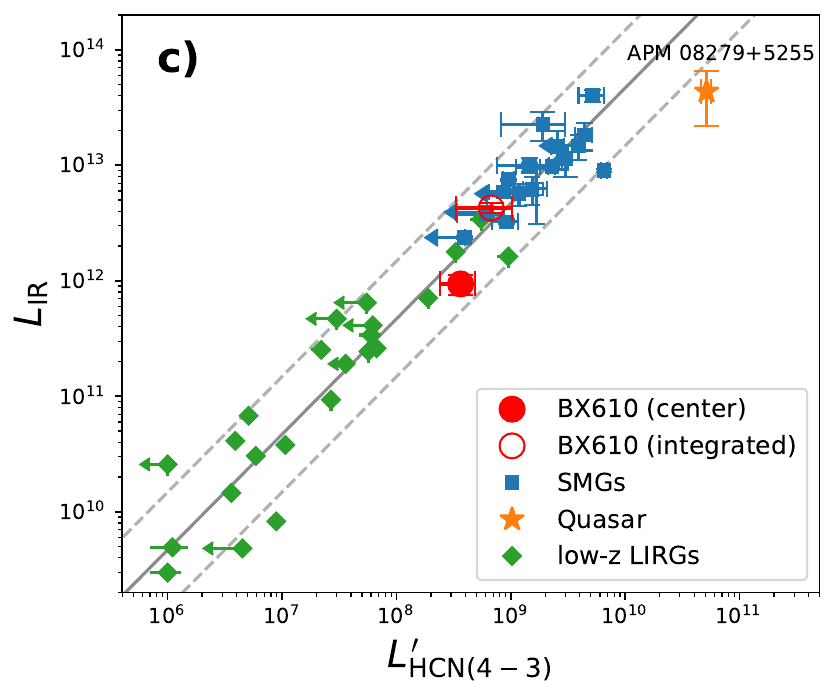}
  \caption{(a) The measured line ratio HNC(5--4)+CN(4--3) versus total IR luminosity, including measurements from lensed SMGs and quasars \citep{Guelin2007,Spilker2014,Bethermin2018,Yang2023}. (b) The derived HCN(4--3)/CO(4--3) ratio versus IR luminosity, including measurements from high-$z$ lensed SMGs and quasars \citep{Spilker2014,Bethermin2018,Yang2023,Rybak2026} and nearby LIRGs \citep{Zhang2014}. (c) The position of BX610 on the dense gas star-formation law between IR luminosity and HCN(4--3) luminosity, as derived for nearby LIRGs \citep{Zhang2014}, together with measurements of lensed SMGs and quasars \citep{Bethermin2018,Yang2023,Rybak2026}. In general, the dense gas line ratios at the centre of BX610 are similar to those of SMGs.}
  \label{fig:HNC_FIR}
\end{figure*}

\begin{figure}[htpb]
  \centering
  \includegraphics[width=0.5\textwidth]{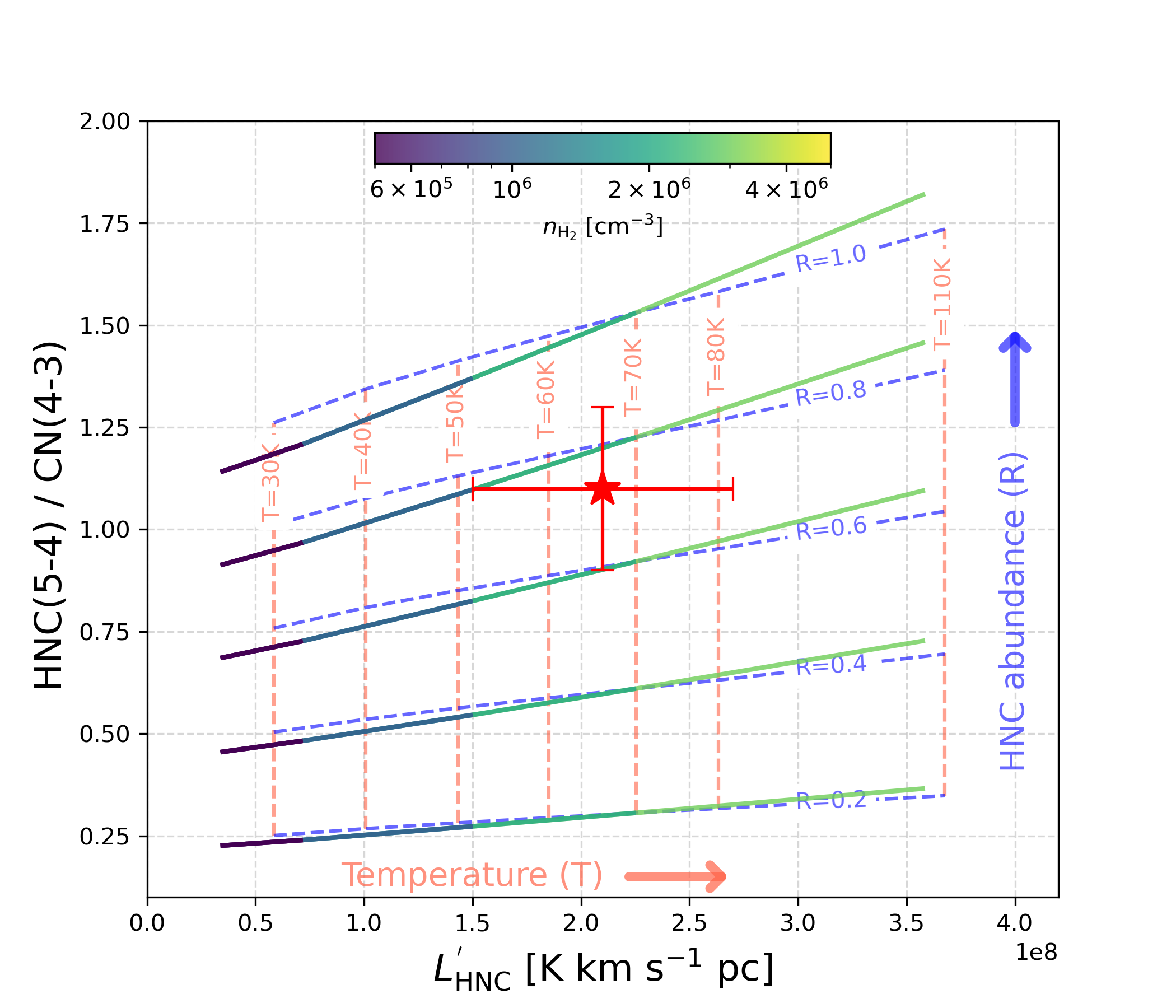}
  \caption{RADEX grids overlaid on the ratio between $N$(HNC(5--4))/$N$(CN(4--3)) and the HNC(4--3) line luminosity. The grids show the parameter space spanned by different kinetic temperatures and different $N$(HNC)/$N$(CN) abundance ratio ($R$) at fixed HNC column density ($1\times10^{14}$~cm$^{-2}$) and molecular gas density ($n_{\text{H}_2}=3\times10^6$~cm$^{-3}$). The coloured lines show the grids at fixed kinetic temperature ($T=65$~K) but with varying molecular gas density ($n_{\text{H}_2}$). There is a strong degeneracy between gas density and temperature. Around $T\sim65$\,K, the observations favour the presence of dense gas with a density of about $3\times10^6$~cm$^{-3}$.}
  \label{fig:cn_hnc_ratio}
\end{figure}

\subsection{Line emission}

We first compared the spatial extent of different cold gas tracers, including CO(4--3) and the dust continuum at rest-frame 650\,$\mu$m, in Fig.~\ref{fig:maps}.
As noted in \citet{Arriagada-Neira2025}, CO(4--3) in BX610 reveals an extended molecular gas disk, while the dust is more concentrated in the galactic centre.
HNC(5--4)+CN(4--3) is even more compact than the dust continuum, largely confined in the centre, with its radial profile following the point spread function (PSF) closely.
To estimate the central compactness of the different components, we computed the flux ratio between the galactic centre ($r<\text{1.2\,kpc}$) and the total integrated flux.
The central region contributes only $21\pm1$\% of the total CO(4--3) emission and $23\pm2$\% of the total dust emission, but it accounts for about $54\pm22$\% of the HNC(5--4)+CN(4--3) emission, though the latter carries a large uncertainty owing to its low S/N.
Nevertheless, the more concentrated distribution of dense gas compared to CO supports a higher dense gas fraction in the galactic centre of BX610.

The full width at half maximum (FWHM) of the decomposed HNC(5--4) and CN(4--3) are similar to that of CO(4--3) in the galactic centre.
However, the decomposed HNC(5--4) and CN(4--3) lines show a small velocity offset of $\sim37\pm26\,\text{km\,s}^{-1}$ relative to CO(4--3) (see also Fig.~\ref{fig:spectra}).
Given the limited S/N, it is difficult to conclude whether this offset is real, but this small offset rules out a fast outflow origin for the dense gas \citep[e.g.][]{Aalto2012,Garcia-Burillo2014}.

\subsection{Line ratios}

We first compared the integrated line ratio of HNC(5--4)+CN(4--3) to CO(4--3) with those of SMGs and quasars spanning a range of IR luminosities.  
This comparison is independent of the line decomposition between HNC(5--4) and CN(4--3), and many literature measurements lack sufficient signal-to-noise for an accurate decomposition.
As shown in Fig.~\ref{fig:HNC_FIR}, the measurement at the centre of BX610 is similar to that of lensed SMGs at comparable redshifts \citep{Spilker2014, Bethermin2018}, while the ratio averaged over the whole galaxy is almost three times lower.
This integrated line ratio further supports the inference drawn from the CO(4--3)/[C\,{\sc i}](1--0) ratio that the centre of BX610 exhibits star-forming conditions very similar to those of compact SMGs \citep{Arriagada-Neira2025}.

Our line decomposition yields an HNC(5--4)/CN(4--3) ratio of $1.05\pm0.23$.
This ratio is similar to that found in the SMG NCv1.143 reported by \citet{Yang2023} and in nearby star-forming regions for their low-J transitions \citep{Aalto2002}, but lower than the ratio of $\sim$1.7 found in the quasar APM 08279+5255 \citep{Wang2011,Yang2023} and the HyLIRG PLCK\_G244.8+4.9 \citep{Canameras2021}.
Both lines trace warm, dense gas.
At a kinetic temperature of $T=65$\,K, CN(4--3) has a critical density of $n_\text{crit}=7.6\times10^{6}$~cm$^{-3}$ and HNC(5--4) is about $n_\text{crit}=1.4\times10^{7}$~cm$^{-3}$ \citep{Shirley2015}.
Theoretically, the HNC/CN ratio can be suppressed in the outer layers of photodissociation regions (PDRs), where CN is enhanced while HCN can be dissociated by strong UV radiation.
In dense, warm environments such as AGN surroundings, however, HNC(5--4) can be boosted by mid-IR pumping or hot-temperature chemistry in X-ray dominated regions \citep{Aalto2007,Meijerink2007}.
Our observations cannot robustly exclude the possible AGN contribution to HNC(5--4) emission, but the low HNC(5--4)+CN(4--3)/CO(4--3) and HNC(5--4)/CN(4--3) ratios disfavour a strong AGN contribution to the HNC emission. Future observations that cover multiple transitions of HNC, CN, and HCN will provide stronger diagnostics.

Assuming that the detected HNC(5--4) is not heavily affected by a buried AGN in BX610, we can use the HNC/CN ratio to constrain the properties of the dense gas.
Since HNC(5--4) and CN(4--3) differ in critical density by only a factor of a few, we can safely model them under a single-cloud assumption.
We used the non-LTE radiative transfer code RADEX \citep{vanderTak2007} with a large velocity gradient (LVG) setup to sample the possible parameter space, together with molecular line data from the Leiden Atomic and Molecular Database (LAMDA) \citep{Schoier2005}.
Specifically, we used the CO, CN, HCN, and HNC data from \citet{Yang2010,Dumouchel2010,Kalugina2015}.
Due to the strong degeneracy among parameters, we fixed the HNC column density to $N({\rm HNC})=1\times10^{14}$~cm$^{-2}$, adopted from the detailed modelling of NCv1.143 by \citet[][]{Yang2023}, which exhibits very similar HNC and CN line flux to BX610.
We also restricted the molecular gas kinetic temperature to the range 30--100~K and the gas density to the range $10^5$ to $10^7$~cm$^{-3}$, which covers most of the parameter space observed in various star-forming regions \citep{Aalto2002,Yang2023}.
The RADEX model grids and our measurement are shown in Fig.~\ref{fig:cn_hnc_ratio}, which favour a gas density of $2-4\times10^6$~cm$^{-3}$ and a temperature of about 50--80~K.
Given our simplified model assumption, this gas condition only reflects the average gas properties of the line-emitting regions.
The strong degeneracy between gas kinetic temperature and molecular gas volume density means that the same observed values can also be reproduced by a lower gas density combined with a higher kinetic temperature, but these values become more extreme as a two times lower density will require nearly twice the temperature, making it far larger than that of nearby star-forming regions.
The model also turns a $N(\text{HNC})/N(\text{CN})$ abundance ratio of $0.7\pm0.1$, indicating that CN is slightly more abundant than HNC in the dense clouds.

The ratio of HCN or HNC to less-dense gas tracers, such as CO, at similar frequencies can serve as a proxy for the warm star-forming dense gas fraction.
Here, we focus primarily on HCN(4--3) and CO(4--3), for which we can compile a range of reasonable literature results spanning a sufficiently broad parameter space.
First, we need to convert the measured HNC(5--4)+CN(4--3) line luminosity to HCN(4--3).
The current observations do not provide sufficient constraints to perform this conversion theoretically, owing to the unknown HCN/HNC abundance ratio and the large allowed parameter space inferred above.
Instead, we adopted an empirical approach and used the stacked spectrum of 22 DSFGs from \citet{Spilker2014}, which yields an average ratio of $L^\prime_\text{HCN(4--3)}/(L^\prime_\text{HNC(5--4)}+L^\prime_\text{CN(4--3)})\approx 0.54\pm0.17$, also similar to the sample-averaged value in four lensed SMGs \citep{Bethermin2018} and NCv1.143 \citep{Yang2023}.
We believe this is a relatively safe approach as the dense gas ratio in the centre of BX610 is very similar to SMGs, as indicated by Fig.~\ref{fig:HNC_FIR}a.
We also applied the same conversion to the galactic integrated values, which are less certain but can be useful for understanding global averaging effects on the dense gas measurements.
Based on this empirical conversion, the derived HCN(4--3)/CO(4--3) ratio is again similar to that of SMGs (see Fig.~\ref{fig:HNC_FIR}b), although with a total IR luminosity that is an order of magnitude lower.
However, when averaged across the whole galaxy, the derived HCN(4--3)/CO(4--3) ratio falls closer to that of nearby LIRGs, highlighting the importance of beam averaging and aperture correction when interpreting dense gas ratios \citep[e.g.][]{Zhang2014}.
On the well-established line luminosity $L^\prime_\text{HCN(4-3)}$ and total IR luminosity $L_\text{IR}$ scaling relation, showing in Fig.~\ref{fig:HNC_FIR}c, both the integrated and central measurements of the BX610 fall closely on the correlation derived from local LIRGs, where the global averaging effect contributes to the scatter of the correlation.

\section{Discussion and conclusions}
\label{sec:discussion}

The detections of both HNC(5--4) and CN(4--3) indicate the presence of dense gas with densities on the order of $2-4\times10^6\,\text{cm}^{-3}$ and kinetic temperatures around 50--80\,K.  
The luminosity ratio of HNC(5--4)/CN(4--3) is near unity, and the RADEX model favours a slightly higher abundance of $\text{N(CN)}$ relative to $\text{N(HNC)}$.  
These suggest that the detected dense gas resides in typical photodissociation regions (PDRs) within starburst regions \citep{Boger2005}, with negligible influence from an AGN, consistent with optical spectral line diagnostics \citep{ForsterSchreiber2014,Newman2014}.  
The different spatial extents between the dense gas traced by HNC(5--4)+CN(4--3) and that traced by CO(4--3) serve as a cautionary note for interpreting the dense gas fraction of similar systems measured from unresolved, lensed configurations. 
These configurations can dilute the dense gas fraction and suffer from differential lensing between different cold gas tracers.  
It is noteworthy that most high-$z$ measurements are galaxy-integrated values from strongly lensed systems, which trace only galaxy-averaged quantities and may be affected by differential lensing.
Since SMGs are predominantly compact starburst systems, a direct comparison to their integrated values may still be valid \citep{Rybak2026}, but future high-resolution follow-up observations are required to verify these studies.

BX610, a representative main-sequence galaxy at cosmic noon, closely follows the scaling relation between $\log L^\prime_\text{dense gas}$ and $\log L_\text{IR}$ derived in the nearby Universe, supporting the view that dense gas mass directly drives SFR at this early cosmic epoch.  
BX610 exhibits a higher molecular gas fraction and a larger star-formation efficiency (SFE) in the central region compared to the outer disk \citep{Arriagada-Neira2025}, consistent with the predominant dense gas in the centre from this study. 
In theory, the higher star formation rate could be driven either by a larger dense gas fraction or by a higher SFE of the dense gas itself.  
With only detections of high-J transitions of HNC and CN, it is difficult to obtain an accurate estimate of the total dense gas mass \cite[e.g.][]{Rybak2026}; thus, it is non-trivial to conclude which process dominates the central starburst.
Meanwhile, as argued by \cite{Jiao2025,Jiao2025a}, the SFR could be primarily determined by the total mass of gravitationally bound gas clouds, which possess a universal dense-gas fraction and SFE \citep[see also][]{Gao2004}.
Nevertheless, the fact that the ratio $L^\prime_\text{HNC(5--4)+CN(4--3)}/L^\prime_\text{CO(4--3)}$ in the nucleus is nearly three times higher than the galaxy-integrated value implies a larger concentration of dense gas in the central region. This may arise either from an intrinsically higher dense-gas fraction or from a larger proportion of the gas mass being confined within gravitationally bound clouds.
Meanwhile, the centre of BX610 follows the near-unity slope between $\log L^\prime_\text{HCN(4-3)}$ and $\log L_\text{IR}$, suggesting that its SFE is at least comparable to that of typical nearby starburst regions and SMGs at similar redshifts, which are found to be much higher than that of Milky Way-like galaxies.

%

Based on nearby kpc-scale resolved studies, both the dense gas fraction and the dense gas SFE have been found to vary across different galactic environments \citep{Garcia-Burillo2012,Usero2015,Gallagher2018,Neumann2025}.  
In several well-resolved sub-kpc studies, the dense gas also exhibits different behaviour in regions affected by spiral arms and bars \citep[e.g.][]{Querejeta2019,Sanchez-Garcia2022,Beslic2021,GU2026}.  
All these studies suggest that the connection between dense gas and star formation also depends on the local dynamical environment of the host galaxy.
Sensitive JWST near-IR images clearly demonstrate the presence of two prominent arms and a possible embedded bar in BX610.  
Based on kinematic analyses of both ionised and cold molecular gas, high-velocity radial gas inflows appear to dominate along the spiral arms (or at the bar ends) \citep{Genzel2023}, supporting efficient radial gas transport through the combined action of spiral arms and/or a bar \citep[see also][]{Ubler2024,Huang2025,Pastras2025,Jolly2026}.  
The detection of dense gas in the centre of BX610 thus corroborates the view that efficient gas transport and the large central gravitational potential can facilitate the condensation of star-forming dense gas at cosmic noon, which naturally leads to an elevated central SFR.

In conclusion, we have reported the first detection of HNC(5--4) and CN(4--3) in a massive, unlensed main-sequence galaxy at cosmic noon, BX610, revealing the concentration of dense gas in its centre.  
Its galaxy-averaged dense gas fraction is similar to that of local star-forming galaxies, while its core region falls closer to that of SMGs.     
Our results are consistent with the view that the SFR is primarily regulated by the availability of the densest gas, which can be enhanced by the efficient cold gas transport in the gas-rich star-forming galaxies at cosmic noon. 
At cosmic noon, both HNC(5--4) and CN(4--3) can be observed simultaneously with CO(4--3) and [C\,I](1--0), which provides an efficient way to probe molecular gas at different densities.  
Future surveys following the same strategy will yield a statistical sample for direct comparison of cold gas at various critical densities.

\begin{acknowledgements}
\label{acknowledgements}
JC is very grateful for the valuable discussions with Zhi-Yu Zhang and Padelis Papadopoulos.
JC, NMFS, LL, GT, JE, CB acknowledge funding by the European Union (ERC Advanced Grant GALPHYS, 101055023).
R.H.-C. thanks the Max Planck Society for support under the Partner Group project "The Baryon Cycle in Galaxies" between the Max Planck for Extraterrestrial Physics and the Universidad de Concepción. R.H-C. also gratefully acknowledge financial support from ANID - MILENIO - NCN2024\_112 and ANID BASAL FB210003.
GM and H\"U acknowledge funding by the European Union (ERC APEX, 101164796). 
Views and opinions expressed are, however, those of the author(s) only and do not necessarily reflect those of the European Union or the European Research Council. Neither the European Union nor the granting authority can be held responsible for them. 
H\"U also thanks the Max Planck Society for support through the Lise Meitner Excellence Program.
This paper makes use of the ALMA data 2013.1.00059.S, 2017.1.01045.S, and 2019.1.01362.S. ALMA is a partnership of ESO (representing its member states), NSF (USA) and NINS (Japan), together with NRC (Canada), MOST and ASIAA (Taiwan), and KASI (Republic of Korea), in cooperation with the Republic of Chile. The Joint ALMA Observatory is operated by ESO, AUI/NRAO and NAOJ. All the data used in this work are publicly available from the ALMA science archive \hyperlink{ALMA Science Archive}{https://almascience.eso.org/alma-data}.
This work is based in part on observations made with the NASA/ESA/CSA James Webb Space Telescope. The data were obtained from the Mikulski Archive for Space Telescopes at the Space Telescope Science Institute, which is operated by the Association of Universities for Research in Astronomy, Inc., under NASA contract NAS 5-03127 for JWST. These observations are associated with program GO 9407.
\end{acknowledgements}

\bibliographystyle{aa} 
\bibliography{dense_gas, dense_gas_preprint} 

\begin{appendix}

\end{appendix}

\end{document}